\documentclass[final,times,twocolumn]{elsarticle}

\journal{Data \& Knowledge Engineering}

\usepackage{url}
\usepackage{amsmath}
\usepackage{amssymb}
\usepackage{graphicx}
\usepackage{csquotes}
\usepackage{enumerate}
\usepackage{booktabs}
\usepackage[T1]{fontenc}
\usepackage{amsthm}
\newtheorem{definition}{Definition}

\usepackage{xcolor}
\newcommand{\predicate}[1]{\texttt{#1}}
\newcommand{\evpredicate}[1]{\predicate{has\_#1}}
\newcommand{\evattrpredicate}[1]{\predicate{has\_attrib\_#1}}
\newcommand{\entity}[1]{\text{#1}}

\let\emptyset\varnothing

\begin{document}
\begin{frontmatter}

\title{A Conceptual Model for Attributions in Event-Centric Knowledge Graphs}

\author[inst1]{Florian Plötzky\corref{cor1}}
\ead{ploetzky@ifis.cs.tu-bs.de}
\cortext[cor1]{Corresponding author}

\affiliation[inst1]{organization={Institute for Information Systems, Technische Universität Braunschweig},
            country={Germany}}

\author[inst2]{Katarina Britz}
\author[inst1]{Wolf-Tilo Balke}

\affiliation[inst2]{organization={CAIR, Department of Information Science, Stellenbosch University},
            country={South Africa}}

\begin{abstract}
The use of narratives as a means of fusing information from knowledge graphs (KGs) into a coherent line of argumentation has been the subject of recent investigation.
Narratives are especially useful in event-centric knowledge graphs in that they provide a means to connect different real-world events and categorize them by well-known narrations.
However, specifically for controversial events, a problem in information fusion arises, namely,  multiple \emph{viewpoints} regarding the validity of certain event aspects, e.g., regarding the role a participant takes in an event, may exist.
Expressing those viewpoints in KGs is challenging because disputed information provided by different viewpoints may introduce \emph{inconsistencies}.
Hence, most KGs only feature a single view on the contained information, hampering the effectiveness of narrative information access.
This paper is an extension of our original work and introduces \emph{attributions}, i.e., parameterized predicates that allow for the representation of facts that are only valid in a specific viewpoint.
For this, we develop a conceptual model that allows for the representation of viewpoint-dependent information. 
As an extension, we enhance the model by a conception of viewpoint-compatibility.
Based on this, we deepen our original deliberations on the model's effects on information fusion and provide additional grounding in the literature.
\end{abstract}




\end{frontmatter}

\section{Introduction}
\label{sec:introduction}
The last couple of years have been marked by several large-scale events.
Beside the global COVID-19 pandemic outbreak, international tensions and war have been a constant topic in the news along with analysis and opinion concerning them. 
Indeed, events of such a scale shape our world, and hence, they are the subject of research from different perspectives, including event prediction \cite{radinsky2012eventprediction,gottschalk2019happening}, event-driven forecasting~\cite{kalifa2022eventforecasting}, and drawing conclusions from historical event analogies \cite{axelrod2017historicalanalogies,ploetzky2021eventanalogies}.
Additionally, research concerning events from ontological \cite{guizzardi2013towardsontologicalevents,almeida2019eventsasentities} and Semantic Web perspectives ~\cite{vanhage2011sem} has gained traction over the last decade.
Those endeavors have led to a better understanding of the nature and composition of events and the construction of structured repositories of event data, i.e.,  Event-Centric Knowledge Graphs (ECKGs).
Such ECKGs, for example EventKG~\cite{gottschalk2018eventkg}, are usually based on the Resource Description Framework (RDF)~\cite{klyne2014rdfprimer} and an underlying schema like the Simple Event Model~\cite{vanhage2011sem}.
Thus, they benefit from a clear conceptual model in their event representation and allow for structured querying and reasoning tasks.

\begin{figure*}
    \centering
    \includegraphics[width=.8\linewidth]{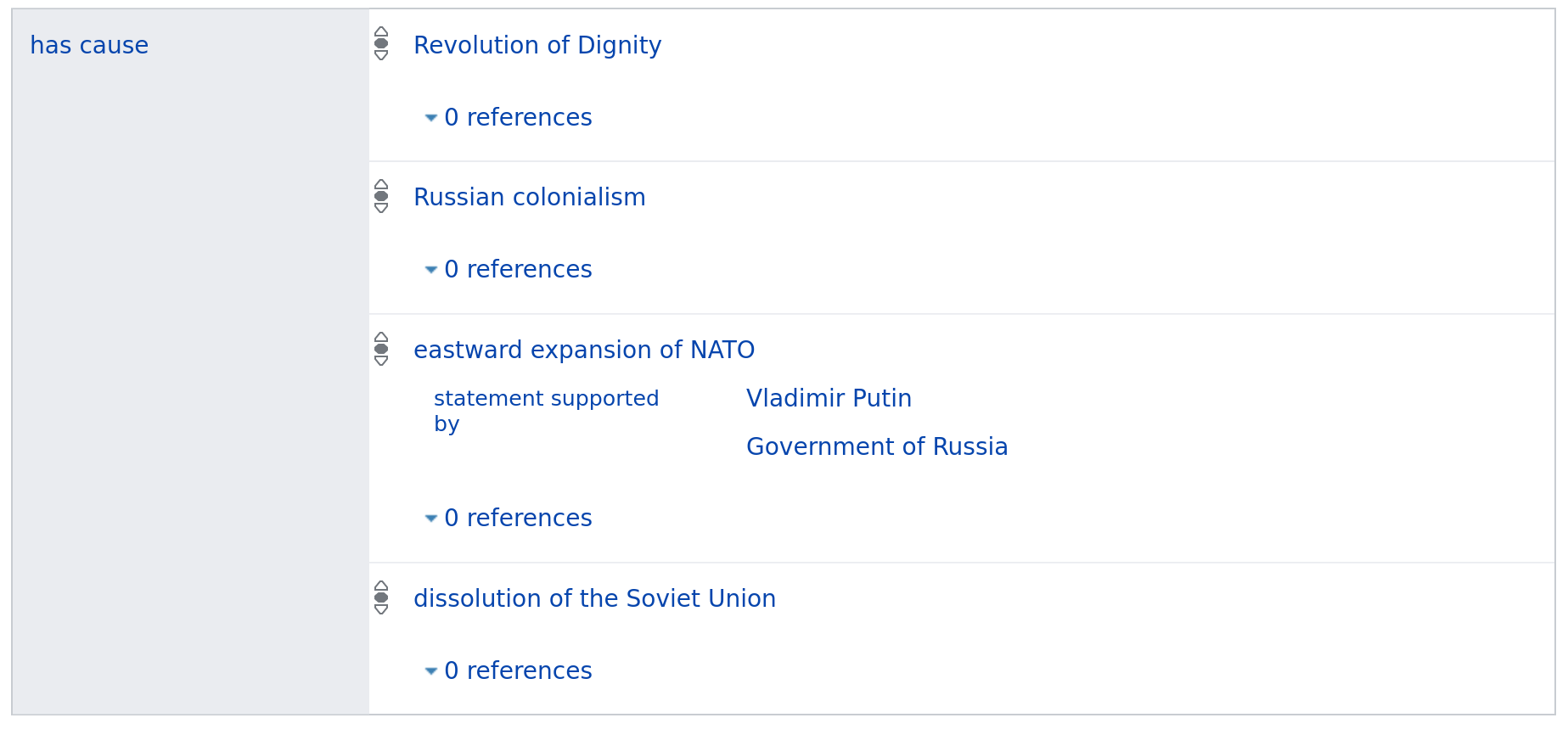}
    \caption{War causes for the Russo-Ukrainian War according to Wikidata (Q15860072).}
    \label{fig:ruvsukr_war_cause}
\end{figure*}
In practical ECKGs such as the aforementioned EventKG and event-centric portions of general purpose KGs (like Wikidata~\cite{vrandecic2014wikidata}), however, there is a lack of a special kind of event information that we called \emph{narrative aspects} in a prior work~\cite{ploetzky2022narrativeaspects}.
That is, factual information, claims and rather subjective stances utilized in describing the event in discourse, e.g., the role a specific participant played in an event, the nature of the event, or the reason an event happened.
Hence, we decoupled events from the narratives surrounding it.

Narratives in general play a big role in understanding and interpreting information \cite{laszlo2008scienceofstories}.
They allow for both an intuitive understanding of knowledge by applying a specific framing of facts \cite{herman2001cognitivenarratives} and at the same time for constructing coherent lines of argumentation, e.g., in strategic communication \cite{wilson2018strategicnarratives}.
In this paper, we follow \cite{kroll2020narratives,ploetzky2024receventsemantics} and understand a narrative in the context of information systems as a recursive graph structure that connects events, event participants (entities), and further descriptions of both (as literals).
Prior works showed how narratives can be constructed from unstructured data \cite{ploetzky2024receventsemantics} and be used as query templates in information systems \cite{kroll2020narratives,kroll2023discoverysystem}, i.e., some parts of a given narrative are marked by variables which are then substituted by a narrative query.

In the context of real-world events, narratives can be used to describe the discourse around a specific event and describe aspects such as narrative frames \cite{frermann2023narrativeframes}, i.e., how specific participants in the event have been perceived.
Oftentimes, such narratives take the form of substitutions for prototypical narrative patterns, i.e., established patterns are isomorphically matched into some bigger knowledge graph, and for each match nodes can be filled with concrete instances (entities or literal values), fleshing out the actual narrative(s). 
Overall and in conjunction with ECKGs, narratives can be seen as a means to make sense of connected subgraphs \cite{kroll2020narratives}.
That is, events might be categorized by well-known narrations, e.g., a conflict might be categorized as an instance of the well-known biblical David vs. Goliath story \cite{ploetzky2022narrativeaspects}.

Unfortunately, categorizing an event in this way based on information from ECKGs leads to the aforementioned problem: real-world ECKGs oftentimes do not provide information on narrative aspects. 
Simply introducing, e.g., narrative frames, to an ECKG, however, may lead to a major problem, namely \emph{inconsistencies} regarding the information needed.
In fact, especially in events that naturally cause controversy (e.g., wars and political conflicts), multiple \emph{viewpoints} concerning a specific property of an event exist.
To illustrate this problem, consider the Russo-Ukrainian War and its representation in Wikidata.\footnote{\url{https://www.wikidata.org/wiki/Q15860072}}
Specifically, the attribute \predicate{has\_cause} as depicted in Fig.~\ref{fig:ruvsukr_war_cause}.
Two of the war causes listed are the \enquote{eastward expansion of NATO} and \enquote{Russian colonialism}.
Those two causes are mutually exclusive when used in a narration since they imply different roles (or narrative frames) of Russia in the conflict.
In the case of a NATO eastward expansion (which boils down to NATO aggression), Russia's role would be a \enquote{defender} from a foreign power.
If one argues for Russian colonialism as the war cause, Russia would rather be seen as a \enquote{conqueror} or \enquote{occupier}, i.e., an aggressive party in the conflict.
Additionally, the first cause is annotated with a \enquote{statement supported by} qualifier, expressing that this property expresses the view of Vladimir Putin and the government of Russia.
The government of Ukraine would presumably argue for the second option, even though it is not stated in Wikidata.

One could argue that there is also some truth in both causes, effectively rendering Russia's role in the conflict to be ambivalent.
We argue, however, that such \emph{viewpoint-dependent information} can not be used arbitrarily.
Instead, they must be contained in their respective viewpoints.
This observation leads to the question: what do these different viewpoints and attributions mean with respect to later information fusions for querying and reasoning?
Matching narratives over information provided from several viewpoints might lead to inconsistent arguments and thus, in the worst case might completely invalidate the explanatory power that should be provided by narrative argumentation patterns.

Introducing different viewpoints in ECKGs is not only useful when dealing with narratives.
The representation and fusion of viewpoint-dependent information is a general problem that was mentioned several times (e.g., \cite{porzel2022narrativizingkgs,hitzler2010reasonablesemanticweb}) but rarely tackled.
Prior works in this direction usually separated viewpoint-dependent or disputed information from the KG by either framing it as an extraction problem~\cite{ploetzky2022narrativeaspects} or as an enrichment of KG data~\cite{porzel2022narrativizingkgs}.
Related work on models for ECKGs mentions the idea of assigning different roles to participants~\cite{vanhage2011sem}, but neither provides a conceptual model on this mechanism nor explains the implications of such roles for downstream tasks.
Hence, the problem of representing viewpoint-dependent information is not solved and rarely tackled beside the benefits, a viewpoint-enabled ECKG would provide especially with respect to fusing information into narratives for plausibility estimations or event categorization.
Therefore, if we would be able to represent information from different viewpoints we could construct narratives from different points of view (analogous to \cite{porzel2022narrativizingkgs}) or verify whether an event follows a given narrative pattern~\cite{ploetzky2022narrativeaspects}.

This paper is an extended version of our previous work presented at the International Conference on Conceptual Modeling (ER) in 2023 \cite{ploetzky2023attributions}. 
In this paper we develop a conceptual model that allows for the representation and fusion of viewpoint-dependent information in ECKGs, and show its strengths and weaknesses in an extensive case study.
Thus, our contributions are fourfold:

\begin{enumerate}[(i)]
    \item We contribute a model for viewpoints on facts in ECKGs. 
    \item We introduce the notion of parameterized predicates in ECKGs that build a taxonomy of KG predicates and link specific instances of predicates to viewpoints. Furthermore, we show how those predicates can be implemented in state-of-the-art ECKGs.
    \item We conduct an extensive case study on a timely, large-scale event and show how the model works along with its limitations.
    \item As an extension to our original work \cite{ploetzky2023attributions} we introduce the concepts of \emph{view\-point-compatibility} and \emph{viewpoint-consistency} for view\-point-en\-abled ECKGs. Furthermore, we deepen our deliberations concerning the consequences of information fusion on viewpoint-consistency.
\end{enumerate}

These contributions are a precursor for effectively working with narratives in ECKGs.
Please note that the latter part will not be the focus of this paper.
Instead, we will focus on introducing viewpoint-dependent information in ECKGs but not on how they can be linked to narratives (this is done elsewhere, especially in \cite{ploetzky2024receventsemantics}).
Additionally, we do not discuss ways to automatically construct viewpoint-enabled ECKGs or automatically derive viewpoints and their relationships.
That is, beside deciding the schematic information (e.g., the predicate labels and associated semantics) the modeler also decides which viewpoints are allowed, how they are connected, and what kind of viewpoint-dependent information can be expressed in the ECKG. 

Before we introduce our model in Sec.~\ref{sec:modeling_viewpoints_in_eckgs}, we discuss its' design principles by reviewing the current representation of the Russian invasion of Ukraine in Wikidata to study the way viewpoint-dependent information is represented in state-of-the-art KGs and the problems of this representation in Sec.~\ref{sec:example}.
Furthermore, we conduct a case study and discussion of the strengths and limitations of our model based on the aforementioned invasion in Sec.~\ref{sec:case_study}.
Finally, we discuss related work in Sec.~\ref{sec:related_work} before concluding the paper in Sec.~\ref{sec:conclusion}.

\section{A Motivating Example}
\label{sec:example}
As a motivation, we revisit the representation of the Russo-Ukrainian War.
More specifically, we focus on the invasion of Ukraine that started in February 2022 and refer to this event as RUvsUKR for the remainder of this paper.
In this section we take a closer look on the Wikidata representation of RUvsUKR\footnote{\url{https://www.wikidata.org/wiki/Q110999040}} in order to study the desiderata for modeling viewpoint-dependent information of events.
Wikidata is not only one of the largest freely available knowledge repositories but also offers a convenient way of adding meta-data to facts, i.e., \emph{qualifiers}.\footnote{\url{https://www.wikidata.org/wiki/Help:Qualifiers}}
In this example, we take a closer look at the usage of such qualifiers for RUvsUKR.

\subsection{Views on Event Attributes}
\label{subsec:views_event_attrib}
\begin{figure*}
    \centering
    \includegraphics[width=.52\linewidth]{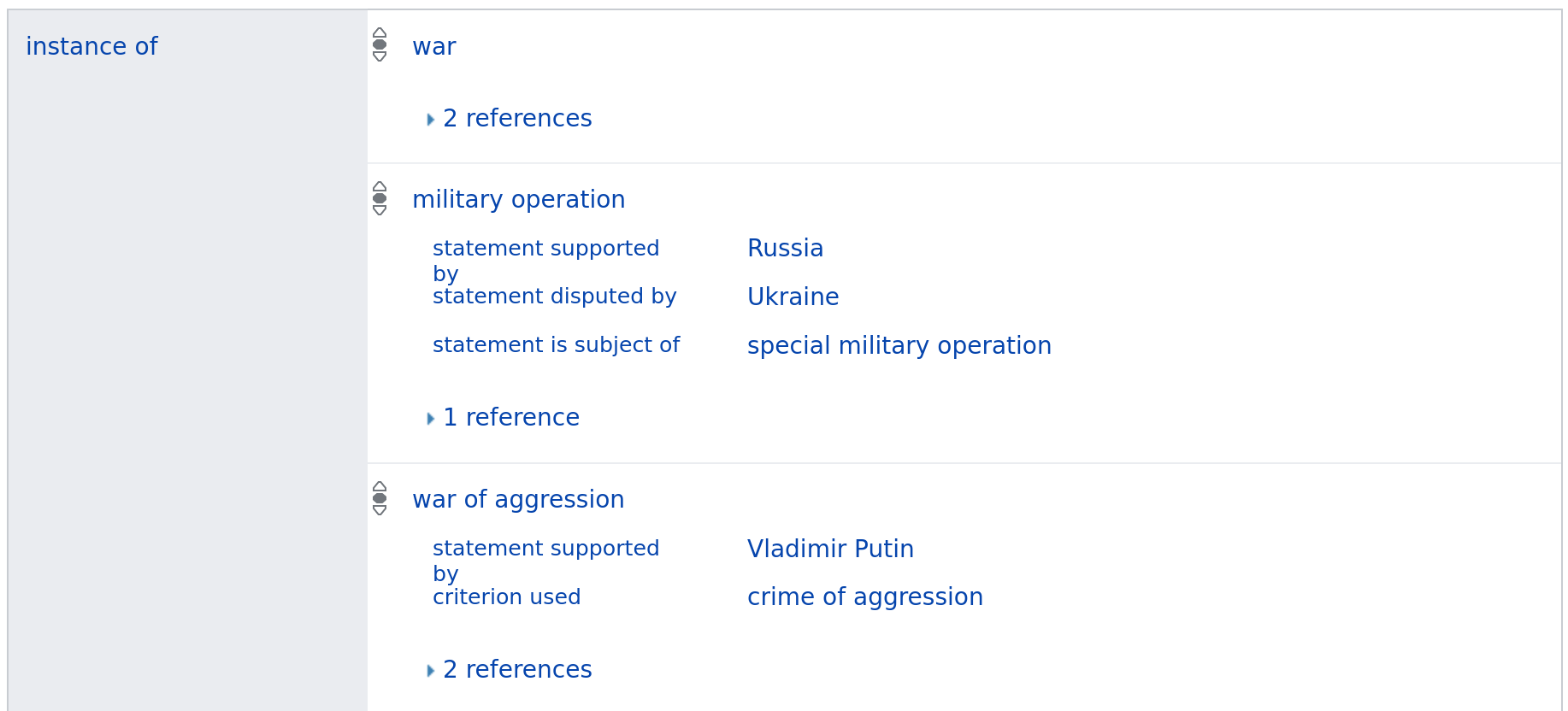}    
    \caption{Excerpt \enquote{instance of} attribute for the Russian invasion of Ukraine in Wikidata (Q110999040). The image was cropped to a number of representative property values.}
    \label{fig:rus_invasion_instance_of}
\end{figure*}

The event is an instance of seven different classes; a selection is depicted in Fig.~\ref{fig:rus_invasion_instance_of}.
One of those classes is \enquote{military operation} and it is qualified as being supported by Russia and disputed by Ukraine.
Additionally, the statement is qualified by a special item that explicitly describes the euphemism \enquote{special military operation} as used by Vladimir Putin to describe the invasion.
This example shows on the one hand that it is disputed whether the event can be seen as a military operation or not.
On the other hand, by referencing the \enquote{special military operation} entity, one can infer that the statement is not actually a military operation but framed as such.

The intention here might be to incorporate Vladimir Putin's framing while at the same time adhering to the notion that the event is in fact an invasion.
That is, at least Vladimir Putin (and possibly the government of Russia) attributes the event in this way; other individuals or groups disagree with this notion.
We can observe this kind of annotation in the same event in multiple cases, e.g., regarding the cause and goal of the event.
However, shoehorning in different views like this leads to modeling and interpretation problems.

For the first problem, we refer again to the disputed event type \enquote{military operation}, which is annotated with a \enquote{statement supported by} and \enquote{statement disputed by} qualifier regarding the respective entities Russia and Ukraine.
However, both entities represent the respective countries in terms of both states.
It is not clear what constitutes this viewpoint, or in other words, which view exactly is the \enquote{Russian} and if it is the view of multiple groups, how is it composed. 
Therefore, the first problem concerns the \emph{viewpoint constitution}.

Taking this into account, the questions arise, which viewpoints are actually important to model, and who are the representatives of the groups behind a viewpoint?
For instance, regarding the goals of the invasion, the view of the head of the Chechen Republic is annotated in Wikidata, but it is not clear whether this is his personal opinion or his opinion as head of state.
Additionally, one might be interested in generalized views like the \enquote{Chechen view} on the conflict.
Is it, however, fair to generalize the view of the head of state to the general view of the whole country?
That is, do we have to define the \emph{viewpoint representatives} that influence a viewpoint?

The last problem concerns practical implications regarding other event attributes, especially the role of participants in the event.
If we adopt the Russian view of framing the event as a military operation, Russia's role significantly differs from its role in an invasion.

\begin{figure*}
    \centering
    \includegraphics[width=.52\linewidth]{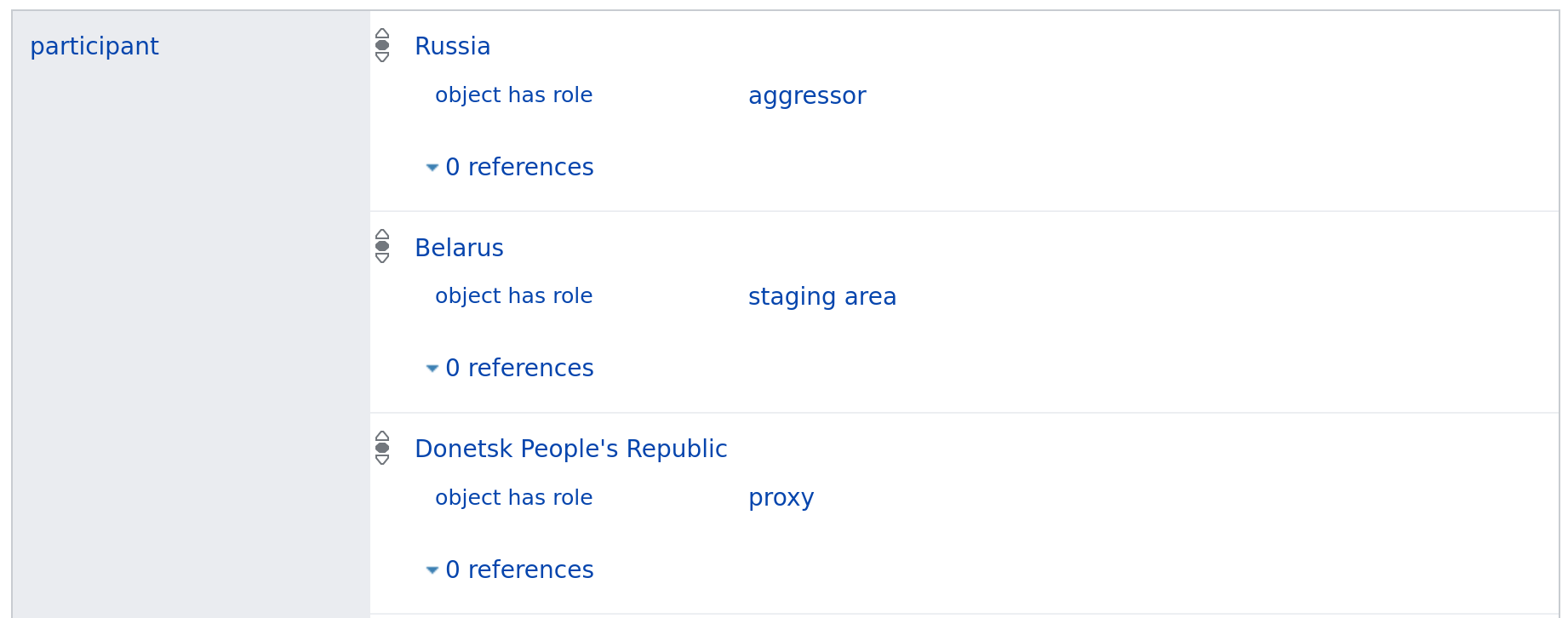}
    \caption{Excerpt from the participants in the Russian invasion of Ukraine according to Wikidata (Q110999040).}
    \label{fig:rus_invasion_participants}
\end{figure*}

\subsection{Views on Roles of Event Participants}
\label{subsec:views_event_partip}
Wikidata supports an \enquote{object has role} qualifier.
In our example, this is used to assign roles to the invasions' participants, e.g., it is used to cast Russia as an aggressor, Ukraine as a war victim, and Belarus as a staging area (an excerpt is depicted in Fig.~\ref{fig:rus_invasion_participants}).
However, as discussed before, this assignment is not universally agreed on.
If we assume that this event is a military operation with the goal of denazification, Ukraine can not be seen as a war victim in the same line of argumentation.
The relation between event participants and events is, however, of high interest and often debated.
Therefore, role assignments should also include the respective viewpoint of the assigner.
This is especially true if we introduce roles that bear moral dimensions, like Russia being a liberator in the event \cite{porzel2022narrativizingkgs}.
We call those assignments \emph{attributions}.
Regarding attributions, again, some questions must be answered for practical use.

Firstly, the set of available attributions must be determined.
This relates to the problem of the choice of event types.
For instance, the role of an aggressor is plausible in wars but not in football matches.
Secondly, attributions may include participant roles that are not typical in knowledge graphs, like the role of an underdog or the aforementioned liberator.
Of course, such new roles are also tailored to specific event types.
While aggressor and liberator are not suitable roles for wars and football games at the same time, the assignment of an underdog is sensible in both contexts.
Additionally, some attributions have a specific relationship with each other.
For instance, a nation that is attributed to be an occupier in one view can not be a liberator in the same event and according to the same view.
This means that the attributions must not contradict themselves from a single point of view.

Finally, it must be clear which statements are actually subject to a viewpoint, i.e., what the actual statement is.
For instance, in Fig.~\ref{fig:rus_invasion_instance_of}, we can see that the invasion is an instance of \enquote{war of aggression} according to Vladimir Putin.
A war of aggression implies that there is an aggressor, and indeed, as depicted in Fig.~\ref{fig:rus_invasion_participants}, Russia is attributed to be the aggressor in the conflict.
However, can we now infer that Vladimir Putin's stance is that this is a war of aggression \emph{and} Russia is the aggressor in it?
In other words, are those two statements compatible and thus valid from the same point of view?
To answer this question, we need a notion of \emph{viewpoint-consistency} with respect to sets of statements. 

\section{Modeling Viewpoint-dependent Information in Event-Centric Knowledge Graphs}
\label{sec:modeling_viewpoints_in_eckgs}
Over the course of this section we develop a conceptual model with respect to the shortcomings identified in the previous section.
In summary, our conceptual model for viewpoint-dependent information in ECKGs must be capable of (i) describing what a viewpoint is and how it is constituted, (ii) how those viewpoints interact with each other, (iii) how they are linked to information in ECKG, and (iv), how those viewpoint-dependent pieces of information can be fused and kept consistent.

\subsection{Viewpoints and Stances}
\label{subsec:views_and_stances}
Generally, we argue that a viewpoint should always represent the view of either an individual or a group of entities toward a given target.
Hence, a group is a finite set of entities constituted by a given criterion, e.g., political parties, a set of newspapers sharing the same political ideology, or interest groups like non-governmental organizations (NGOs).
Along these lines, a target is a \emph{fact} about an event, like the role a participant played or the kind of happening itself.
Therefore, we can define viewpoints as follows:

\begin{definition}[Viewpoints]
    \label{def:viewpoints}
    A \emph{viewpoint} $v \in V$ is a \emph{consensual stance} $s \in \{\text{\emph{valid}},\text{\emph{invalid}}\}$ towards a fact $f \in KG$ expressed by a group $G$.
\end{definition}

The constitution criterion of $G$ is subject to the modeling domain or extraction method (cf., for example, \cite{quraishi2018viewpoints}).
Thus, the modeler needs to decide which groups should be used in her specific model.
However, later in this paper, we will provide some intuition regarding the selection of viewpoints.
It is possible that the group only consists of a single member to allow viewpoints of individual entities.
The target of the viewpoint is a fact $f$, e.g., whether Russia can be seen as an aggressor in RUvsUKR, from an Knowledge Graph $KG$.

In order to determine whether a group $G$ does in fact accept $f$ as valid, we rely on the notion of stance detection.
Stances are usually defined as attitudes, standpoints, and judgments of a speaker regarding a given proposition \cite{aldayel2021stancesurvey}.
In our case, the proposition is the fact in question, and the speaker is the group where the stance, according to \cite{aldayel2021stancesurvey},  expresses agreement, disagreement, or neutrality towards the facts' validity.
We define a function $stance(g, f) \mapsto \{\text{valid}, \text{invalid}\}$ with $g \in G$  and $f$ being the fact in question. 
The function returns whether $g$ sees $f$ as valid or not.
Note that the second case includes both neutrality and disagreement regarding the validity of $f$.
The latter case implies that $stance(g, \lnot f)$ must hold.

Please also note that setting a fact to be invalid in $v$ if the underlying group stance is neutral leads to several implications.
On the one hand, it prevents neutrality from being interpreted as acceptance.
We argue that facts should only be valid in a viewpoint if an actual majority agrees on it.
Especially in groups with a silent majority (i.e., absence of stances for a majority of group members), this would otherwise lead to viewpoints where facts are valid although being agreed upon by a minority only.
On the other hand, this decision weakens negative stances.
This is because the reason for a negative stance is ignored, and hence, the notion of disagreement of the group proxied by the viewpoint is lost.
In this paper, we argue that the benefit of combining the neutral and negative stances is greater than having the ability to differentiate between them.

To construct a viewpoint for a group we need the individual stances of each group member towards $f$.
This implies that every $g \in G$ must be capable of expressing such a stance.
Additionally, we defined any $v \in V$ to be a \emph{consensual} stance of the group and hence, we need a measure for consensus.
For this paper, we define consensus as:

\begin{definition}[Consensus]
    \label{def:consensus}
    A \emph{consensus} in a group $G$ is reached, if a \emph{weighted consensus measure}  $\phi_W(G, f)$ with weights $W$, surpasses a given threshold $\theta$.
\end{definition}

The choice of consensus measure depends on the domain it is used in.
A typical choice would be a simple majority vote surpassing a threshold $\theta = 0.5$.
However, rigorous measures like a necessary majority of two-thirds are also possible.
Since $\phi$ is a weighted consensus measure it allows us to boost certain individuals in $G$, e.g., in cases where a dedicated speaker or otherwise higher-ranked individual exists.
For instance, in the case of the US senate, it might be suitable to assign a higher weight to the respective party speaker.

With those preliminaries we can define the stance of a group as:
$$
  stance(f, G, W) =  \left\{
    \begin{array}{ll}
        \text{valid} & \,\phi_W(G, f) \geq \theta, \\
        \text{invalid} & \, \text{otherwise}. \\
    \end{array}
\right. 
$$

If the stance regarding $f$ is valid for $G$, we say that \emph{$f$ is valid in $v$}, where $v$ is the viewpoint representing $G$.  

\subsection{Viewpoint Hierarchies}
\label{subsec:view_hierarchies}

Until now, we have a notion of viewpoints and consensus to construct a viewpoint for a given group.
Different viewpoints are, however, not generally disjoint but can be combined in various cases. 
A fact $f$ might be valid in two viewpoints $v_1, v_2 \in V$, i.e., there are two groups with a positive stance towards $f$.
Hence, those viewpoints are compatible with respect to $f$ and can be \emph{aggregated} into a new viewpoint $v^*$.
In other words, $v^*$ subsumes $v_1$ and $v_2$ with respect to $f$, and only with respect to $f$.
For other facts, aggregation might not be possible.
However, research in the area of viewpoint discovery has shown that similar groups show similar stances on the same topic, e.g., groups constituted by homophily \cite{quraishi2018viewpoints}.
That means that we can define typical groups in the sense of common aggregations between groups to prevent a combinatorial explosion of viewpoint aggregations.
For that, we define a viewpoint hierarchy as follows:

\begin{definition}[Viewpoint Hierarchy]
    \label{def:viewpoint_hierarchy}
    A \emph{viewpoint hierarchy} is a directed acyclic graph $H(N, A)$ with $N = V \cup \{\text{\emph{ALL}}\}$ being a set of viewpoints and $A$ a set of arcs indicating \emph{aggregation} relationships.
\end{definition}

\begin{figure}[t]
    \centering
    \includegraphics[width=\linewidth]{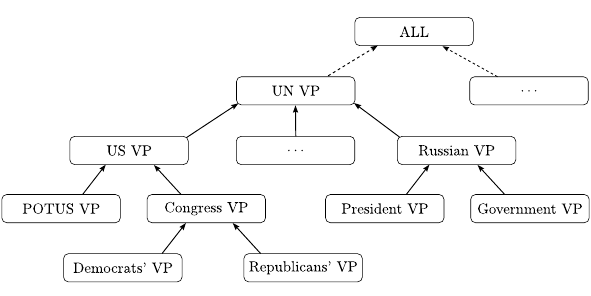}
    \caption{An example for a viewpoint aggregation hierarchy based on our motivational example for a fact $f$. For the abbreviations: US means United States, UN means United Nations, and POTUS means President of the United States.}
    \label{fig:viewpoint_hierarchy}
\end{figure}

Such a hierarchy describes typical aggregations of viewpoints, an example is depicted in Fig.~\ref{fig:viewpoint_hierarchy}.
As a concrete example: the aggregation between the US Congress, Democrats, and Republicans indicate that a fact valid for Democrats and Republicans is typically also valid for the US Congress.
The shape of the hierarchy depends on the usage domain, e.g., viewpoint hierarchies for political issues will include other viewpoints than those tailored for sports events.
Regardless of the shape, we set ALL to be the \emph{virtual top} of the hierarchy.
If a fact is valid in ALL, it is also valid in all other viewpoints of the hierarchy.
In other words, a fact valid in ALL is said to be universally agreed upon.
In all other cases, a fact is only valid in some viewpoints.

Arranging viewpoints in a hierarchy has practical advantages.
On the one hand, the hierarchy defines which groups and thus viewpoints should actually be represented and which are implicitly co-represented.
On the other hand, such a hierarchy exemplifies which information from a knowledge graph can be fused without introducing inconsistencies.

However, besides defining the shape of the hierarchy, the main problem is the definition of suitable aggregation functions and consensus measures between the viewpoints.
A weighted consensus measure, analogous to Def.~\ref{def:consensus}, can be used in this respect.
That is, while subsuming $n$ viewpoints concerning a fact $f$ to a larger viewpoint $v^*$ each individual viewpoint might have a different weight on the decision whether $f$ is valid in $v^*$.
Thus, using such a hierarchy also requires a choice on how to handle non-agreeing viewpoints during aggregation.
For instance, assume two viewpoints $v_1, v_2$ that can be subsumed to $v^*$. 
If $v_1$ has a higher weight than $v_2$, a fact might be valid for $v_1$ and due to the higher weight also in $v^*$ but not for $v_2$.
Therefore, inconsistencies in the hierarchy may appear and $v^*$ can not be seen as a representative of all parts it is aggregated from.
We propose two hierarchy variants to approach this problem.

\paragraph{View-preserving Hierarchies (VPH)} We allow for dissenting viewpoints in the (sub) hierarchy.
Hence, if any fact $f$ is valid in a viewpoint $v$ it need not be valid in all sub-viewpoints of $v$.
The benefit of this approach is that minority viewpoints can exist, especially in larger hierarchies.
Typically, democratic systems work by majority votes, and hence, it is inevitable that a dissenting minority exists.
VPH allows us to preserve those minorities' stances on $f$.
Note, however, that if $f$ is valid in ALL it must also be valid in all other viewpoints, i.e., for those facts, no minorities exist.

\paragraph{Winner takes all Hierarchies (WTAH)}
This variant assumes that after aggregation the fact in question is valid in all viewpoints of the subtree.
In other words, if, for instance, viewpoints $v_1, v_2, v_3$ are subsumed by $v^*$ any fact valid in $v^*$ is also valid in $v_1, v_2,$ and $v_3$ regardless of their individual stance on the fact.
This leads to benefits in practicability of downstream tasks since it reduces the overall complexity of the model.
The major advantage of WTAH over VPH is the simple handling facts since it is possible to safely combine facts from different subtrees if the fact is valid in the roots of those subtrees.
Also in practical terms, WTAH is quite often used to form common opinions, e.g., in political party systems.
However, like all winner takes all systems, this approach might lead to  a representation paradox in larger hierarchies where a fact is valid in $v^*$ that would never be valid in one specific sub-viewpoint $v$ of $v^*$ because it might for instance, be harmful or nonsensical for this fact to be valid in $v$.
We discuss this problem again in the case study in Sec.~\ref{sec:case_study}.

Finally, we can now generalize our notion of a viewpoint.
If we assume a viewpoint hierarchy as defined in Def.~\ref{def:viewpoint_hierarchy}, we also have a notion of typical groups in our model.
It is very likely that each group will have a stance on multiple facts.
In this case, a viewpoint with respect to a group is a \enquote{world view} that describes all facts assumed valid or invalid of that group.
More formally:

\begin{definition}[Viewpoints (Generalization)]
    \label{def:viewpoints_wrt_groups}
    Given a viewpoint hierarchy $H$, a set of facts $F \subseteq KG$ and a group $G$.
    A \emph{viewpoint} (with respect to $G$) is a consensual stance $s \in \{\text{\emph{valid}}, \text{\emph{invalid}}\}$ for each $f \in F$ with respect to $H$.
\end{definition}

As an effect, each fact $f \in F$ that is valid in ALL is also valid in $v$.
The validity for all facts that are not valid in all depends on the aforementioned hierarchy variants for $H$.
Please note that there might be some facts in a KG for which we do not know whether they are valid in $v$, hence $F \subset KG$ might be very well the case in practical scenarios.
Finally, for the case $|F| = 1$ the generalized definition of viewpoints is identical to Def.~\ref{def:viewpoints}.

Until here we discussed desideratas (i) and (ii) for our conceptual model. 
Before we discuss (iii), i.e., how to link viewpoints to ECKGs, we first define our notion of events and ECKGs in the following.

\subsection{Events and Event-Centric Knowledge Graphs}
ECKGs have gained traction over the last years, either by means of constructing specialized KGs \cite{gottschalk2018eventkg,gottschalk2021openeventkg} or by using portions of general knowledge graphs \cite{rudnik2019newseventkg}.
Events themselves have been studied extensively from an ontological perspective \cite{almeida2019eventsasentities,guizzardi2013towardsontologicalevents,scherp2009feventmodel} and from a semantic web perspective \cite{vanhage2011sem}.
Mostly agreed upon here is the notion that an event has a temporal as well as a spatial component and connects participants in a certain situation.
Also, a notion of hierarchy is often described, i.e., the aggregation of single events to form complex events.
For our purposes in this paper, we rely on the following simple definition:

\begin{definition}[Events]
    Let $P$ be a set of entities and $T$ be a set of literals referring to points in time. An \emph{event} is an interaction between participants $p \in P$ that takes place at a specific point in time $t \in T$ or in a specific time span $(t_1, t_2) \in T \times T,\,\,\textrm{with}\,\, t_1 < t_2$. 
   We denote the set of events as $EV$. 
\end{definition}

The definition includes the aforementioned typical components of events.
Each event must at least provide attributes for specifying the time and location it takes place in.
We do not specify the granularity of time and space, since for our purposes the existence is enough.
Further properties might exist but are not mandatory for our purposes.

At the core, each event describes an interaction between participants, the latter being entities.
Typically, the event \emph{label} describes this interaction, as is the case for practical ECKGs.
Additionally, we assign an \emph{event type} to each event, denoted by $event\_type(ev)$ with $ev \in EV$.
With that, we can formally define ECKGs.

\begin{definition}[Event-Centric Knowledge Graphs]
    An \emph{event-centric know\-ledge graph} is a \emph{knowledge graph}, represented by RDF triples $\langle s, p, o \rangle$ (sub\-ject-pre\-di\-cate-ob\-ject), where each subject is an event.
\end{definition}

\begin{figure}[t]
    \centering
    \includegraphics[width=\linewidth]{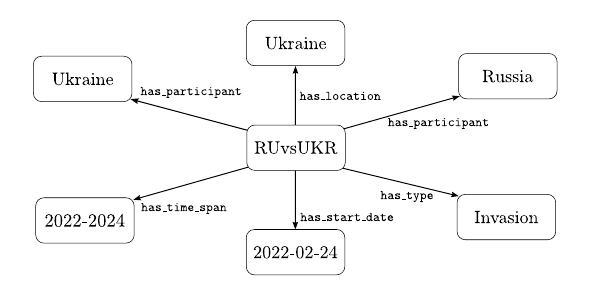}
    \caption{Example ECKG for an excerpt of the motivational example.}
    \label{fig:eckg}
\end{figure}
In other words, ECKGs are star-shaped knowledge graphs with events at their centers.
A knowledge graph here is nothing more than a graph following the RDF standard recommendations \cite{klyne2014rdfprimer}.
An example graph for illustration purposes is depicted in Fig.~\ref{fig:eckg}.
Events may also be present in a triple's object, e.g., for sub-event relationships.
Subjects, however, must be events. 
Predicates in ECKG either denote attributes of an event (e.g., the time and location) or connect entities as participants to the event. 
In the most general case, this is done with a \evpredicate{participant} predicate.
Objects can either be entities, events, time, locations, or literal values.

\subsection{Predicate Parameterization and Refinements}
\label{subsec:refinement}

Refering back to our motivational example, we can observe that the relationships between events and participants are a key part of ECKGs.
Such relationships, like participant roles or other forms of further event characterization, are expressed by predicates in the graph that also bear specific semantics.
Additionally, predicates in an ECKG follow a specific ontology and can oftentimes, analogous to viewpoints, described by specialization hierarchies.
For instance, while each participant of an event can be connected by a \evpredicate{participant} predicate to it, the event type can allow for more specialized predicates like \evpredicate{war\_party} for wars or \evpredicate{election\_candidate} for elections.
We call the concept of specializing predicates according to the event type a \emph{refinement}.
To indicate that a predicate in a triple can be refined, we denote \emph{et} as a parameter to it, i.e., the triple becomes $\langle \entity{s}, \text{\texttt{p}}_{et}, \entity{o} \rangle$.
Given the predicate vocabulary $\mathcal{P}$ of an ECKG, the parameterized predicates, with respect to the event type, are a proper subset $\mathcal{P}^{ET} \subset \mathcal{P}$.
For the \enquote{regular} predicates $\mathcal{P}^{REG} \subset \mathcal{P}$, i.e., predicates that can be applied to all event types, we set $\mathcal{P}^{ET} \cap \mathcal{P}^{REG} = \emptyset$.
Hence, any regular predicate can not be used as a parameterized predicate.

\begin{figure*}[t]
    \centering
    \includegraphics[width=.75\linewidth]{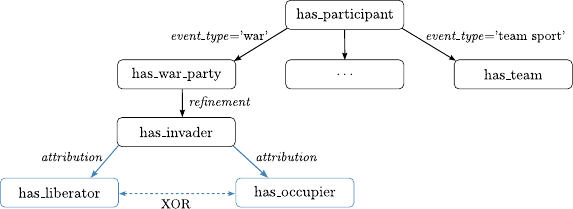}
    \caption{Example for a predicate hierarchy featuring a specialization operation and mutual exclusiveness constraint (XOR). The upper black-colored portion depicts an example of parameterized predicates and their refinements as introduced in Sec.~\ref{subsec:refinement}. The blue-colored lower portion depicts attributions as introduced in Sec.~\ref{subsec:attributions}.}
    \label{fig:predicate_taxonomy}
\end{figure*}

We additionally allow for multiple refinements.
For instance, a war party may also have the role of an invader for one participant.
In this case, we may refine \evpredicate{war\_party} to \evpredicate{invader}.
Therefore, the predicates can be arranged in a hierarchy. Fig.~\ref{fig:predicate_taxonomy} illustrates an example predicate hierarchy in its upper, black-colored portion.

In accordance to the refinement, the event type may also impose constraints on the former.
For instance, an event with type \enquote{invasion} should have at least two participants.
One of them should be an invader, and the other one should be an invaded country. 
Additionally, invaders and invaded countries can not be the same participants.

\subsection{Attributions}
\label{subsec:attributions}
In this section we discuss desiderata (iii), i.e., how we can link viewpoints to triples in ECKGs.
As a precursor, in the last section we introduced parameterized predicates to indicate possible refinements based on the event type.
Those refinements are helpful to add precisifications to event-related facts, i.e., they specialize the relationships between the events and their attributes and participants.
However, in the motivational example, we have seen a second class of those precisifications, i.e., disputed or viewpoint-dependent information.
This is especially true for event-participant relationships that incorporate moral dimensions like being a \enquote{liberator} in an invasion instead of a \enquote{occupier}.
Also, disputed information like RUvsUKR being a military operation belongs to this class that is characterized to include facts only valid in a specific viewpoint.
We call those facts \emph{attributions} and define them as follows:

\begin{definition}[Attributions]
       \label{def:attributions}
     An \emph{attribution} is a \emph{parameterized predicate} with a parameter $v \in V$, i.e., $\langle \entity{s}, \text{\texttt{p}}_v, \entity{o} \rangle$ has attribution $\text{\texttt{p}}_v$. We call a triple that contains an attribution a \emph{claim}.
\end{definition}

At this point, we combine our deliberations on viewpoints and viewpoint hierarchies from the previous sections.
Attributions link triples from an ECKG to the aforementioned viewpoint model.
Here, $v$ represents a label that describes $G$, e.g., the NATO or Russia.
Attributions thus connect triples in an ECKG to a given viewpoint hierarchy.

While parameterized predicates, as introduced in Sec.~\ref{subsec:refinement}, are solely dependent on the event type, attributions depend on a viewpoint.
Attributions allow us to introduce a set of predicates $\mathcal{P}^{ATT} \subset \mathcal{P}$ with $(\mathcal{P}^{ET} \cap \mathcal{P}^{REG}) \cap \mathcal{P}^{ATT} = \emptyset$ that describes viewpoint-dependent information.
Again, we can utilize the structure of ECKGs and define a set of \emph{permissible viewpoints} $V^{ET} \subseteq V$ based on the event type.
The intuition is that, in general, only a subset of viewpoints is sensible for certain event types.
A viewpoint hierarchy sensible for the domain of international politics including the viewpoints of different countries and NGOs might not be sensible for the domain of soccer.

Attributions regarding participant roles can be described as further refinements, i.e., as a specialization of a parameterized predicate.
This case is illustrated in blue color in Fig.~\ref{fig:predicate_taxonomy}.
After two refinements, \evpredicate{invader} can be further specialized in the attributions $\evpredicate{liberator}_v$ and $\evpredicate{occupier}_v$.
Both attributions add a moral judgment to the role of an invader.
The first attribution bears a justification for the invasion, the second one condemns the event.
Note that those attributions are mutually exclusive with respect to a viewpoint.
That is, it is not possible to describe an invader both as liberator and occupier from the same viewpoint. 
Therefore, a set of constraints can apply to $\mathcal{P}^{ATT}$. 
We suggest two kinds of constraints:

\begin{description}
    \item[Mutual exclusiveness]  
    This constraint applies to two attributions that can not be applied to the same participant in the same viewpoint.
    For instance, one can not be \enquote{liberator} and \enquote{occupier} at the same time from a certain point of view.
    \item[Inverse role enforcement] This constraint introduces pairs of attributions that are mutually exclusive but always co-occur.
    Hence, if a participant is assigned to an attribution with inverse role enforcement, another participant is attributed to the counterpart.
    This is, for example, the case for \enquote{underdog} and \enquote{topdog} attributions.
    If a participant is attributed as \enquote{underdog} in a conflict, the other participant is automatically attributed as \enquote{topdog}.
\end{description}

Attributions regarding attributes of an event are described as \emph{transformations} of regular predicates. 
If, for instance, the cause of an event is disputed, we might want to have an attribution with the same semantics as \evpredicate{cause}.
In such cases, we transform the \evpredicate{cause} predicate into \evattrpredicate{cause}.
That is, we apply a transformation function that maps a regular predicate $p \in \mathcal{P}^{REG}$ to an attribution $p' \in \mathcal{P}^{ATT}$ while preserving the semantics of $p$.

\begin{figure*}[t]
    \centering
    \includegraphics[width=.65\linewidth]{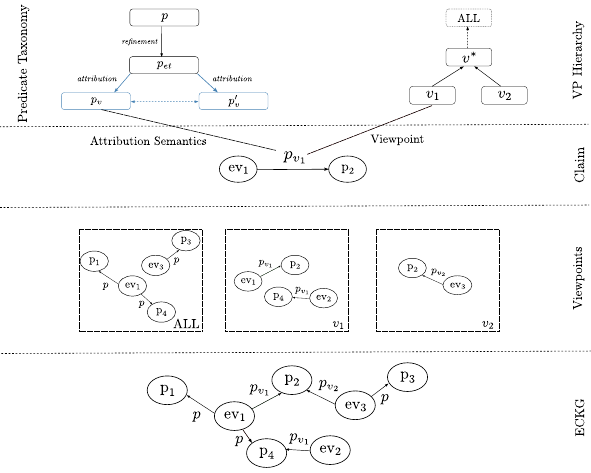}
    \caption{An overview of all concepts introduced in Sec.~\ref{sec:modeling_viewpoints_in_eckgs} and their relationships. An ECKG can be decomposed into viewpoints for different groups (here: ALL, $v_1$, and $v_2$) where ALL denotes all undisputed facts that are valid in all viewpoints. All triples that are not valid in ALL are called claims and marked with an attribution in their predicate that stems from a predicate taxonomy and links to a viewpoint in a viewpoint hierarchy.}
    \label{fig:conceptual_model}
\end{figure*}

One special case, however, is a transformation of the \evpredicate{event\_type} predicate.
Such a transformation could be suitable if the event type itself is subject to debate, like in our motivational example.
Those transformations would have an impact on all other attributions and also on parameterized predicates.
Given that RUvsUKR is a military operation and not an invasion, attributions like an \evpredicate{aggressor} may not be defined for this event type.
In those cases, according to the viewpoint assuming RUvsUKR to be a military operation, all attributions that are not defined in the attribution vocabulary for it are invalid.
The opposite direction is also true.
According to all remaining viewpoints, attributions only defined for military operations are invalid since, according to those viewpoints, the event was an invasion.

\subsection{Claims and Viewpoint-Compatibility}
\label{subsec:viewpoint_compatibility}
The final desiderata (iv) for our model concerns the combination of view\-point-dependent information and the consistency of such combinations of fusions.
In order to discriminate facts in an ECKG containing a regular predicate or parameterized predicate from facts containing an attribution as predicate, we introduced \emph{claims} in Def.~\ref{def:attributions}.
More formally, any fact $f$ from an ECKG (i.e., $f$ is a triple $\langle s, p, o \rangle$) is a claim if $p \in \mathcal{P}^{ATT}$.
As a consequence, each ECKG that allows for attributions can be decomposed into a set of facts $\mathcal{F}$ and a set of claims $\mathcal{C}$.
Naturally, both sets are disjoint.
A visualization of all concepts up to this point is depicted in Fig.~\ref{fig:conceptual_model}.

Considering a viewpoint hierarchy with a set of viewpoints $V$ as nodes, this means that all triples $f \in \mathcal{F}$ are valid for all $v\in V$.
The claim set $\mathcal{C}$ can further be decomposed by their attributions, i.e., for any $v \in V$ we have a set $\mathcal{C}_v$  containing the claims with either $v$ in their attribution parameter or those that can be derived by the hierarchy variant (e.g., in winner takes all hierarchies (WTAH) $\mathcal{C}_v$ will also contain all triples from higher level viewpoints).
From that, we can construct the triple set for viewpoint $v$ according to Def.~\ref{def:viewpoints_wrt_groups} as $\mathcal{T}_v = \mathcal{F}\cup \mathcal{C}_v$.

Each claim is only valid in the viewpoint of its respective attribution and hence, the ECKG stays consistent on a global level.
However, viewpoints must also be self-consistent, i.e., for all $v \in V$ the respective $\mathcal{C}_v$ must be consistent.
Typically, consistency can be modeled by imposing constraints like we have done in Sec.~\ref{subsec:attributions} for attributions.
From our example in Fig.~\ref{fig:predicate_taxonomy} one can derive that 
$$
\mathcal{C}_v = \{\langle \entity{s}, \evpredicate{liberator}_v, \entity{o} \rangle, \langle  s,\evpredicate{occupier}_v,o\rangle\}
$$
for any $v \in V$ is inconsistent.
Such combinations naturally occur during querying ECKGs or reasoning tasks involving multiple rules.
Combinations of facts and claims are called \emph{fusions} during the remainder of the paper.
We call two claims that can be fused without causing an inconsistency in a viewpoint \emph{viewpoint-compatible} and define this property as follows.

\begin{definition}[Viewpoint-Compatibility]
    \label{def:viewpoint_compatibility}
   Two claims $c_1, c_2$ with attributions $p^1_v, p^2_v \in \mathcal{P}^{ATT}$ are \emph{viewpoint-compatible} in $v$ if their fusion is \emph{non-con\-tra\-dic\-to\-ry}.
\end{definition}

We denote viewpoint-compatible claims as $c_1 \simeq_c c_2$ through the remainder of this paper.
Therefore, $c_1 \not\simeq_c c_2$ denotes that $c_1$ and $c_2$ can not be fused without causing an inconsistency.
Note that this definition only concerns claims and not facts since the latter do not contain an attribution.
Hence, facts are valid in all viewpoints, and any fusion solely between facts or facts and, at most, one claim is always viewpoint-compatible.

Determining whether $c_1 \simeq_c c_2$ holds means to check whether the claims contradict themselves, which would render the fusion inconsistent.
In practice, one would rather check for $c_1 \not\simeq_c c_2$.
We outline two general approaches in this regard:

\begin{description}
    \item[Attribution Constraints] 
    We can utilize the aforementioned attribution constraints, i.e., mutual exclusiveness and inverse role enforcement.
    If two claims contain the same subject, different but mutual exclusive attributions with the same viewpoint and the same object, the claims can not be viewpoint-compatible.
    This rule would render our aforementioned example concerning the \evpredicate{liberator} incompatible since $o$ cannot be a liberator and occupier at the same time in the same event.
    \item[Explicit Consistency Rules]
    It is also possible to define explicit consistency rules involving additional knowledge from the ECKG.
    Such rules may be subject to different event types, e.g., a specific rule may only apply to conflict events, another one only to elections.
    Since attributions may oftentimes be derived from refined predicates (as illustrated in Fig.~\ref{fig:predicate_taxonomy}), deriving the applicable set of rules is rather straightforward.
    Defining those rule sets is, again, subject to the modeler analogous to the viewpoint hierarchy.
\end{description}

\subsection{Viewpoint Consistency}
\label{subsec:viewpoint_consistency}
In the previous section, we introduced viewpoint compatibility of claims.
This concept is important in the context of graph construction and maintenance. 
Each triple that is inserted into the graph must not lead to an inconsistency, i.e., viewpoint-compatibility must be ensured for all claims in any viewpoint of a viewpoint hierarchy.
The same goes for queries, i.e., each query result should not contain triples that are viewpoint-incompatible.
For both cases, we introduce the concept of \emph{viewpoint consistency} in this section.

As depicted in Fig.~\ref{fig:conceptual_model} a viewpoint-enabled ECKG can basically be decomposed into general facts $\mathcal{F}$ (that is all triples valid in ALL or its predicate not having an attribution) and different viewpoints $v_1,\dots, v_i \in V$, i.e., a set of claims with the same viewpoint based on a viewpoint hierarchy, denoted by $\mathcal{C}_{v_i}$.
Viewpoint-consistency can now be formally defined as follows:

\begin{definition}[Viewpoint-Consistency]
    \label{def:viewpoint_consistency}
   Given a viewpoint hierarchy $H(N, A)$, a viewpoint $v \in N$, an ECKG $KG$ and a set of claims $\mathcal{C}_v$ stemming from $KG$. 
   We call $\mathcal{C}_v$ \emph{consistent} if $c_1 \simeq_c c_2$ for all $c_1, c_2 \in \mathcal{C}_v$.
   If for all $v \in N$ the corresponding $\mathcal{C}_v$ is consistent, we call $KG$ \emph{viewpoint-consistent}.
\end{definition}

In other words, if we insert a viewpoint according to Def.~\ref{def:viewpoints} in the graph we would assume that all other viewpoints uttered by the group that took a stance on a triple should be compatible to the first one.
Since the general viewpoint of the group according to Def.~\ref{def:viewpoints_wrt_groups} is captured by a claim set in an ECKG, each pairs of claims must be viewpoint-compatible.\footnote{Of course this is an idealized view on how groups find consensus on facts or beliefs in the real world. Actual groups may be struggling to be consistent in their views, especially over time -- the latter factor is not taken into account in this paper and subject to future work.}

In terms of data management in the ECKG, each insertion, update, and deletion must keep the graph viewpoint-consistent.
Please note that insertions (and updates and deletions) of facts preserve viewpoint-consistency automatically in our model since viewpoint-compatibility is only defined for claim fusions in Def.~\ref{def:viewpoint_compatibility}.
Finally, we have to discriminate between the hierarchy variants since both have different implications on which claims are considered for the consistency assessment.

\paragraph{WTAH} In a winner takes all hierarchy, each claim valid in a viewpoint $v^*$ is also valid in all of its sub-viewpoints.
As an effect, if we insert a claim to a sub-viewpoint $v$ of $v^*$ the claim set $\mathcal{C}_{v} \cup \mathcal{C}_{v^*}$ must be consistent in order to keep the graph viewpoint-consistent.
In practice this means that for each claim we have to check viewpoint-compatibilty not only for the viewpoint of its attribution but also in all viewpoints that are located higher up the hierarchy.
Additionally, if we insert a new claim into a higher level of the hierarchy, it is \enquote{pushed down} the hierarchy, and hence, all claims in lower hierarchies that contradict this claim are rendered invalid, thus leading to the corresponding viewpoints to be inconsistent. 
Such scenarios require a cascading update policy to retain consistency.

\paragraph{VPH} The objective of a view-preserving hierarchy is to allow for dissenting viewpoints while still having the advantage of viewpoint aggregation.
In terms of inserting new claims, this means that it is enough for an inserted claim to retain consistency in the respective viewpoint.

\section{A Case Study based on the Russian Invasion of Ukraine}
\label{sec:case_study}
In this section, we conduct a case study based on our motivational example.
Each part of this section considers one modeling task that has to be done in order to apply the model in practice.
That is, we discuss which groups to consider, how consensus is constructed, and how a viewpoint hierarchy can be constructed in Sec.~\ref{subsec:cs_constructing_a_vp_model}.
We further discuss how viewpoint-enabled ECKGs can be constructed with respect to our elaborations concerning parameterized predicates and viewpoint consistency in Sec.~\ref{subsec:cs_constructing_a_vp_eckg}.
Finally, we discuss reasoning about viewpoints in such graphs in Sec.~\ref{subsec:cs_reasoning_viewpoints} before summing up the discussion and an outlook on future work in Sec.~\ref{subsec:cs_discussion}.
In each part, we will provide practical examples for the respective component and discuss alternations and limitations, respectively.

 \begin{figure*}[t]
    \centering
    \includegraphics[width=\linewidth]{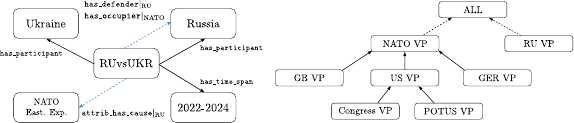}
    \caption{Example ECKG with attributions and a corresponding viewpoint hierarchy.}
    \label{fig:implementation}
\end{figure*}
\begin{figure*}
    \centering
    \includegraphics[width=\linewidth]{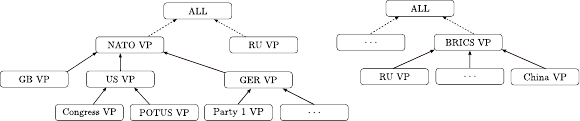}
    \caption{An alternative to the viewpoint model depicted in Fig.~\ref{fig:implementation}.}
    \label{fig:alternative_implementation}
\end{figure*}
 \subsection{Constructing a Viewpoint Model}
 \label{subsec:cs_constructing_a_vp_model}
In this part of our case study we focus on how viewpoint hierarchies are constructed, how viewpoints are handled in terms of the hierarchy, and our design decision concerning neutral stances.

\subsubsection*{Defining Viewpoints in Hierarchies}
The first step in constructing a viewpoint-enabled ECKG is to define the viewpoint hierarchy.
For that, we need to define all viewpoints that are useful for our application scenario.
Since our motivational example concerns RUvsUKR and thus international tension, the viewpoints of governments and NGOs are naturally interesting on the conflict.
The modeling we propose is not limited to organizational viewpoints.
One may also consider the views of individuals, schools of thought in research, and rather conceptual groups like \enquote{the political right} or \enquote{the political left}.

In real life applications (i) the usage domain mostly influences those choices, and (ii), the data that is available also shapes those choices.
The latter questions also influence the viewpoint constitution, i.e., who the group members that can take a stance to different facts and actually create viewpoints on event information are.
In this case study, we use the viewpoint hierarchy depicted in the right portion of Fig.~\ref{fig:implementation}.
Each viewpoint describes a political entity that took a stance RUvsUKR in the past, namely Great Britain (GB), Germany (GER), the NATO, the United States (US), US Congress, the President of the United States (POTUS), and Russia (RU).
The aggregation is based on the real-life composition of those viewpoints.
For example: it is likely that the NATO stance on condemning Russia for RUvsUKR is also taken by all sub-viewpoints.
However, such hierarchies may evolve over time either by extending it with additional viewpoints or splitting existing ones into smaller groups.
Both options are shown in Fig.~\ref{fig:alternative_implementation}.

Such transformations are possible but come with price.
If we only consider a single fact like Def.~\ref{def:viewpoints} does, changing the hierarchy layout is straightforward.
For instance, if a major party in the GER viewpoint would change their view on a claim such as \enquote{Russia's invasion of Ukraine is self-defense}, it might be sensible to split GER into the party the remaining parties and assume VPH to capture this specific minority stance (due to the higher weight of all other parties,  RUvsUKR is not an act of self-defense would still be valid in GER).
However, we introduced the notion of hierarchies specifically as a means for typical aggregations.
If used in an ECKG, changing the layout leads to a re-assessment of viewpoint-consistency for the whole graph.
The model is, hence, more suitable for scenarios with well-defined groups that are known beforehand, even though changing the model is possible but costly.

\subsubsection*{Viewpoint variants}
Assume the statement \enquote{Russia is the aggressor in RUvsUKR} or as a triple $\langle \entity{RUvsUKR}, \evpredicate{aggressor}, \entity{Russia}\rangle$ and the viewpoint model in Fig.~\ref{fig:implementation}.
The obvious first step is to determine whether different stances in the groups shared by the viewpoints exist.
If this is the case, the viewpoint hierarchy and variant decide what happens with the remaining viewpoints.
For this study, assume that the statement is valid in Congress, POTUS, and NATO. 
In addition, we know that it is not valid in RU.
We now have to discriminate between VPH and WTAH.

\paragraph{WTAH} Since the statement is valid for Congress and POTUS it is by aggregation also valid in US.
Even if the statement would not be valid in one of the aforementioned, due to WTAH, it is valid in US since it is valid in NATO.
The same is true for GB and GER, hence the statement is valid in both viewpoints.
For RU we stated that it is not valid.

\paragraph{VPH}
The statement is valid in Congress, POTUS, and NATO. 
By aggregation, it is also valid in US.
We do, however, not know the viewpoints for GER and GB in this model.
This has implications on the downstream tasks which will be discussed in the remainder of the case study.

At this point we also point out a representation paradox that may occur in larger hierarchies like the one depicted in Fig.~\ref{fig:viewpoint_hierarchy}.
Assume the fact \enquote{Russia is a war criminal in RUvsUKR} and the example hierarchy as depicted in Fig.~\ref{fig:viewpoint_hierarchy}.
Obviously, this fact would not be valid in the Russian viewpoint.
If, however, it is valid in enough viewpoints on the same hierarchy level, due to the aggregation it would be valid in the UN viewpoint and hence, in the whole subtree including Russia.
Of course, this effect is by design but would lead to a situation where Russia was a war criminal according to the Russian view.
One can easily observe that this would hamper information fusion from a Russian point of view.
A solution approach to mitigate this problem could be boosting the weights according to a fact or increasing the consensus threshold in higher levels of the hierarchy.
The first option could, for example, be implemented in a way, where viewpoints of groups represented in the fact in question are boosted.
In the end, a trade-off between a plurality of viewpoints and the complexity of using the hierarchy in downstream tasks prevails.
It is up to the domain, which configuration should be used.
We argue that both variants provide a solid ground for the representation of viewpoints.

\subsubsection*{Considerations and Limitations}
Concerning the viewpoint model we discuss the following considerations and limitations:

\paragraph{Choice of hierarchy variant} The viewpoint hierarchy and respective variant must be known in advance.
Like with all explicit models, this requirement can be hard to fulfill.
More research in the area of (semi-)automatic construction of such hierarchies is necessary to either assist modelers with this task or automate it completely.

\paragraph{Static perspective on viewpoints} Another problem concerning the viewpoint hierarchy is the rather static understanding of viewpoints and opinions.
In reality opinions may change over time, especially for controversial events.
Time is, however, not a factor in the current model.
By restructuring the viewpoint hierarchies there is an option to explore this possibility further in future work.
Additionally and finally, the model does not capture any influences between viewpoints beside a one-dimensional aggregation.
In our example hierarchies (Fig.~\ref{fig:viewpoint_hierarchy} and Fig.~\ref{fig:implementation}) the view of the President of the United States will most likely have an impact on Congress and Senate members, especially those in the same political party as the president.

\paragraph{Neutral stances are not represented}
Especially for VPH, in our example, we end up with the following explanations for the absence of viewpoints for $\langle \entity{RUvsUKR}, \evpredicate{aggressor}, \entity{Russia}\rangle$ in GER and GB:
\begin{enumerate}
    \item The groups behind GER and GB explicitly stated that they remain neutral in their stance towards Russia being the aggressor in RUvsUKR.
    \item We do not have enough data for the case.
    \item There was no consensus either because of failing to reach one or too many neutral stances since neutral stances during consensus build are interpreted as negative stance (cf. Sec.~\ref{subsec:views_and_stances}).
\end{enumerate}
Since a viewpoint can only express a positive or negative stance, we can not discriminate between the three cases mentioned.
While introducing a \enquote{true neutral stance} would be a possibility, this would lead to more complex downstream tasks.
Still, the absence of an explicit neutral stance might be a limiting factor for some application scenarios.

\subsection{Constructing a Viewpoint-enabled ECKG}
\label{subsec:cs_constructing_a_vp_eckg}

Having defined a viewpoint hierarchy the next steps concerns the construction of a viewpoint-enabled ECKG.
In the last years, the construction and curation of knowledge graphs have been a constant and well-researched topic (cf. \cite{weikum2021curationknowledgegraphs}).
One of the core parts is defining the vocabulary for the graph.
In our model we also have to define the parameterizations, i.e., what refinements and attributions we want to allow in our graph.
This includes both, attributions concerning potentially disputed information and transformed regular predicates.
Again, this process is highly domain-dependent.
For our example we utilize the graph in the left portion of Fig.~\ref{fig:implementation}.
Here, \evpredicate{participant} and \evpredicate{time\_span} are regular predicates for which we assume no disputes.
This need not be the case for all domains.
For instance, there might be disputes on the time span for longer running events or if the start date is disputed.
Our attributions encompass \evpredicate{defender} and \evpredicate{occupier} as examples for refinements for \evpredicate{participant}, and \predicate{attrib\_has\_cause} for a transformed regular attribute (\evpredicate{cause}).

\subsubsection*{Materialization of claims}
Parameterizing predicates to represent attributions allows for direct integration in knowledge graphs.
That means that both facts and claims can co-exist in the same graph.
We rely on RDF reification~\cite{schreiber2014rdfprimer} to represent attributions in RDF.
Specifically, we utilize \emph{singleton properties}~\cite{nguyen2014singeltonproperty} as a reification technique in this example.
At the core, singleton properties model reification with the idea of representing one specific relationship between two entities.
This can be seen analogously to the idea of an attribution that is specific to one viewpoint.
Additionally, singleton properties generate fewer triples than standard RDF reification and have been shown to represent reified knowledge as well as other techniques like named graphs in a large-scale study on Wikidata \cite{hernandez2015reificationwikidata}.

Without loss of generality, consider the $\evpredicate{occupier}_{\text{NATO}}$ attribution from Fig.~\ref{fig:implementation}.
We can materialize this claim as follows:
{\small
\begin{align*}
\langle \entity{RUvsUKR}, \predicate{has\_occupier\#1}, \entity{Russia} \rangle\\
\langle \entity{has\_occupier\#1}, \predicate{singleton\_property\_of}, \entity{has\_occupier} \rangle\\
\langle \entity{has\_occupier\#1}, \predicate{acc\_to\_vp}, \entity{NATO} \rangle\\
\end{align*}
}
Note that in this case the constraint that an ECKG only contains triples with an event as the subject is slightly relaxed to allow also reified attributes.
Additionally, both attributions and transformed regular predicates are reified in the same way.
Hence, dividing those concepts does not increase the implementation complexity.

The amount of materialization required is again dependent on the viewpoint hierarchy variant.
For instance, WTAHs only require the materialization of the highest common viewpoints since the triple is also valid for all subtrees of those viewpoints.
Considering the viewpoint hierarchy in the left portion of Fig.~\ref{fig:implementation}, both attributions only require one respective reification.
If the example claim is materialized as shown above, it is clear from the viewpoint hierarchy that it is also valid in the whole subtree, i.e., including the viewpoints GB, US, GER, and transitively US Congress and POTUS.
Additionally, if a claim is valid in all viewpoints on the highest hierarchy level, i.e., in ALL, no materialization is necessary.

Materializing claims for the same hierarchy in VPHs, however, requires a different approach.
We reduced the viewpoints to binary stances and thus, as an implication, we have to materialize the claim for all viewpoints in which they are valid.

However, due to the open world assumption (OWA) that is typically assumed in KGs, the absence of a claim can not be interpreted as a negative claim.
Since a viewpoint can also express a negative stance towards a fact, however, viewpoint-enabled ECKGs must provide a way to represent those negative stances.
Assume $\langle \entity{RUvsUKR}, \evpredicate{attacker}, \entity{Russia}\rangle$.
If we want to express that this claim is not valid in RU we can either utilize predicate transformation for negative predicates, i.e.,  
$$
\langle \entity{RUvsUKR}, \evpredicate{not\_attacker}_{\text{RU}}, \entity{Russia}\rangle
$$
or, in cases where we defined explicit constraints in our predicate taxonomy, by inserting the opposite claim, i.e., 
$$
\langle \entity{RUvsUKR}, \evpredicate{defender}_{\text{RU}}, \entity{Russia}\rangle
$$
The latter case, however, can not be generalized.
If we assume that NATO is a participant in RUvsUKR, Russia not being the attacker does not automatically render Russia the defender in the conflict without additional background knowledge.

\subsubsection*{Mining attributions for automatic ECKG construction}
One task during the construction of viewpoint-enabled ECKGs is attribution mining or attribution extraction.
Since attributions are defined as parameterized predicates, relation extraction techniques can be applied (cf. \cite{detroja2023relationextractionsurvey} for a recent survey).
However, as outlined in our previous work \cite{ploetzky2022narrativeaspects} the automatic detection and extraction of attributions leads to some unique challenges.
For instance, in Fig.~\ref{fig:rus_invasion_instance_of} the \enquote{war of aggression} attribute in Wikidata has a reference that provides provenance information, namely an excerpt from a speech of Vladimir Putin.
Quote: 
\emph{\enquote{Some of our colleagues here have mentioned that they are shocked by “Russia’s ongoing aggression in Ukraine.” Indeed, military operations are always a tragedy for specific people, specific families, and the country as a whole. And we must certainly think about how to stop this tragedy.}}\footnote{\url{http://en.kremlin.ru/events/president/news/72790}, last accessed Nov.~25, 2024}.
Here, Putin refers to Russia's aggression as a quote from members of the G20.
While this snippet can in fact be used to attribute RUvsUKR as war of aggression, it is hard to determine whether Russia is really meant to be an aggressor from Vladimir Putin's point of view.
Most likely this is not the case since Putin defends the war efforts but subtleties in this regard hamper an automatic attribution mining.

\subsubsection*{Considerations and Limitations}

With regard to (automatically) constructing viewpoint-enabled ECKGs, we identify the following limitations for real world applications:

\paragraph{Requirement of same conceptual views} As discussed before, this model is designed to allow for representing viewpoint-dependent \emph{validity} of facts.
This requires, however, the same conceptual understanding of the attributions in all viewpoints.
In other words, inserting an attribution like $\evpredicate{liberator}_{v}$ requires a shared conceptualization of a \enquote{liberator} for all $v \in V$.
While this might be the case for most parts, this might be a problem for extraction algorithms, since primary sources sometimes utilize framing techniques to present the facts in a specific context.
Thus, it must be ensured during fact extraction that the attribution semantics is indeed guaranteed.

\paragraph{Rarity of positive stances} One design decision concerned the fusion of neutral and negative stances to imply the invalidity of a fact.
We already argued why this is beneficial for this model.
However, this again constrains the extraction process, since in reality, a lot of viewpoints may stay neutral.
In this case, the model can be slightly tuned in three ways.
Firstly, it is possible to define finer-grained hierarchies to adapt for extraction sparsity.
Secondly, the weights of the consensus measures might be adapted in a way that allows for faster consensus.
The neutral stance regarding a fact from a viewpoint could also be interpreted as agreement and hence, the fact be treated as valid in this viewpoint.
This option comes with all benefits and flaws we already discussed.
Finally, another possibility for future work would be to explicitly model the neutral case as a third option.
This would of course lead to difficulties in reasoning of such models.
It would, however, allow us to discriminate willful neutrality (for instance not taking any stance on the role of Russia in RUvsUKR) from the absence of a stance.

 \subsection{Viewpoint-consistency and downstream tasks}
 \label{subsec:cs_reasoning_viewpoints}
A viewpoint-enabled ECKG should always be viewpoint-consistent, i.e., the claims of all viewpoints must be consistent.
For that each insertion of a new claim in the KG must not contradict any other claim in the same viewpoint.
Inserting the claim $\langle \entity{RUvsUKR}, \evpredicate{attacker}_{\text{RU}}, \entity{Russia} \rangle$ into the ECKG in Fig.~\ref{fig:implementation} would render RU inconsistent.
Additionally, results of queries must also be viewpoint-consistent, i.e., the fusion of facts and claims in queries must be consistent.

To determine whether two claims $c_1, c_2$ can be fused we introduced the compatibility relation $c_1 \simeq_c c_2$.
The remaining question we tackle in the following is, how to determine which claims we have to compare in this regard when inserting new claims or posing a query to the graph.
This question can be answered by utilizing the viewpoint hierarchy.
For that, assume the following set of claims:
$$
 EKG = \{ c_1|_{\text{ALL}},\, c_2|_{\text{US}},\, c_3|_{\text{POTUS}},\, c_4|_{\text{RU}},\, c_5|_{\text{RU}},\, c_6|_{\text{Congress}}\}
$$
and the viewpoint hierarchy in Fig.~\ref{fig:implementation}.
Each claim in $EKG$ is annotated with the respective viewpoint of its attribution.
We can now revisit the two viewpoint hierarchy variants and discuss which claims can be fused in a query in a viewpoint compatible way.

\paragraph{WTAH}
In a winner-takes-all hierarchy, a claim is valid in all viewpoints of a subtree with root $v*$, if it is valid in $v*$.
Hence, in our example, each claim that is valid in US is also valid in Congress and POTUS.
For $EKG$ this means that $c_2$ is valid in US, Congress, and POTUS.
The other way around, no claim $c|_{\text{US}}$ can exist with $c \not\simeq_c c_3$.

This leads to the observation that we can safely fuse claims in  subtree as long as they follow a path from the root of the subtree.
That is, regardless which attribution is contained in $c_2$, it can always be fused with $c_3$ and $c_6$ because the latter two must be viewpoint-compatible to $c_2$ because US subsumes both POTUS and Congress.
However, two claims whose viewpoints are siblings in the hierarchy (like POTUS and Congress) need not be viewpoint-compatible.
For instance, if the attributions of $c_3$ and $c_6$ are $\evpredicate{liberator}_{\text{POTUS}}$ and $\evpredicate{occupier}_{\text{Congress}}$ respectively they would not be viewpoint-compatible.

In the latter case no consensus would be reached between the two viewpoints and hence we do not know whether claim is valid in US.
Hence, a query engine needs to explicitly check for viewpoint-compatibility if siblings in a hierarchy should be fused.
This is also the case for all viewpoints that do not have a common parent at all like $c_5$ and $c_3$.
However, as long as queries involve fusions in a downward direction of the hierarchy, fusions are simplified and no additional checks for viewpoint-compatibility is needed.

\paragraph{VPH}
In a viewpoint-preserving hierarchy, claims can be invalid in a viewpoint $v$ even if they are valid in an aggregated viewpoint containing $v$.
This is beneficial to preserve minority viewpoints and does simplify inserting and updating facts in the ECKG.
However, this also implies that each fusion of claims in queries needs a check for viewpoint-compatibility regardless of their respective viewpoint.

In general, WTAH and VPH introduce a trade-off similar to the materialization vs. query-time computation trade-off in databases.
While ECKGs using a WTAH are harder to maintain since it is harder to ensure consistency in all viewpoints, querying such graphs might be easier since the viewpoint-compatibility check can be omitted in the aforementioned cases.
In contrast, inserting and updating claims in VPH is straightforward but querying such graphs may need additional compatibility checks.
In the end, the choice of variant depends on the actual use case, but a designer should be aware of this trade-off.

\subsection{Discussion and Future Work}
 \label{subsec:cs_discussion}
Attributions can be used to include disputed or otherwise viewpoint-dependent information, i.e., claims, in ECKGs.
We argue that they can solve portions of the inconsistency problem arising, especially for controversial events, as shown in the motivational example.
The proposed model of attributions along with a viewpoint hierarchy allows for such viewpoint-dependent information to co-exist in the same KG while combining information is guided by the hierarchy without enforcing global consistency.
Additionally, the model can be implemented by only using established Semantic Web technologies, e.g., singleton properties for reification and SHACL~\cite{kublauch2017shacl} to ensure the attribution constraints like mutual exclusiveness.
By now, however, the model is not validated in practice, i.e., an implementation in a real-world application is missing.

Concerning an actual implementation our model provides semantics on how predicates in a graph can bear different viewpoints by attributions.
In addition, the viewpoint hierarchy naturally guides the aggregation of different claims and is a basis for a query semantics.
Such queries can then construct different views on event knowledge that is internally consistent.
Potential applications are plenty, especially for domains with rather clear-cut groups that oftentimes share different opinions.

Concerning limitations of the model, we already provided the main limitations over the course of this section. 
For us, the major downside concerns its lack of dynamic elements, especially time.
Opinions may change fast and hence, there might be limitations towards building larger knowledge graph for domains that require rapid updates of knowledge.
More work especially in deploying this works for multiple domains is necessary to verify its scalability in this regard.

\section{Related Work}
\label{sec:related_work}
\paragraph{Viewpoint discovery} 
Research in the area of discovery and description of viewpoints is mostly done concerning document collections or social media (e.g., \cite{hada2023echochambers}) or by learning public opinion with language models \cite{chu2023bertmediadiet}.
Discovering viewpoints on controversial topics in social media, for example, has been done by clustering users based on interaction graphs \cite{quraishi2018viewpoints} or by applying variants of topic modeling \cite{thonet2017viewslda}.
Additionally, advances in stance detection \cite{aldayel2021stancesurvey} and a growing understanding of the difference between stance and sentiment \cite{bestvater2023sentimentvsstance} may soon allow for more fine-grained methods of collecting viewpoints from text corpora.
Our work contributes to this area by providing a means to represent those different viewpoints by using established formats for downstream applications.

\paragraph{Viewpoints in ECKGs}
Current ECKGs, either constructed from general-pur\-pose knowledge graphs \cite{gottschalk2018eventkg,gottschalk2021openeventkg} or from news \cite{rospocher2016newseventkg}, do not support different views on facts.
Their underlying RDF schema, the simple event model (SEM)~\cite{vanhage2011sem}, however, intend the use of different views but does not describe in detail how those views work in terms of their composition or implications for downstream tasks.
Other works have suggested fusing factual knowledge from KGs with viewpoint-dependent knowledge from other sources in downstream tasks either at query time by using a hybrid query-processor~\cite{ploetzky2022narrativeaspects} or by designing transformation pipelines on top of a graph query~\cite{porzel2022narrativizingkgs} or by constructing narratives and linking them to ECKGs~\cite{ploetzky2024receventsemantics}.

\paragraph{Multiple viewpoints in Ontologies and Conceptual Modeling}
Related areas of conceptual modeling, like requirements engineering and enterprise modeling, have adopted notions of multiple viewpoints in the past \cite{kotonya1996reqviewpoints,sommerville1999processinconsistencyviewpoints,sultan2018w6hframework}.
In contrast to our work, those views limit the model to certain aspects that are relevant in it but do not modify the underlying facts or assumptions based on the view taken.
More similar to our work are approaches of ontology integration~\cite{osman2021ontologyintegration} and specialized description logics allowing for reasoning with different ontology vocabularies (e.g., \cite{alvarez2022standpointlogics,hemam2011mvpowl,stuckenschmidt2006multiviews}).
The difference of those works to our approach is that information fusion in our model seeks sufficient agreement between different viewpoints, in order to aggregate knowledge graphs representing the different viewpoints. 
The graph structure representing all the facts from a specific viewpoint can be viewed as an input ontology, but instead of matching equivalences between entities from different input ontologies in order to create an ontology alignment which then guides the ontology integration, the viewpoint hierarchy is used to guide the aggregation of viewpoint-compatible facts.

\section{Conclusion}
\label{sec:conclusion}
Overall, attributions based on viewpoint hierarchies can improve the utility of ECKGs.
On the one hand, they allow for the representation of disputed and morally charged information.
Both kinds of knowledge would otherwise either not be available in an ECKG or reasoning tasks in this regard would suffer since fusing information in a meaningful way is only possible with a clear conceptual understanding.
On the other hand, all conceptualizations shown in this paper can be implemented by using already available techniques orchestrated by the models developed here.
However, future work is necessary to put this model into practice, e.g., to actually enrich available ECKGs with attributions and test the practicability at scale.
The development of efficient information fusion algorithms and methods to automatically construct viewpoint-enabled ECKGs is subject to future work on this topic.
Additional work on the model itself is possible in terms of interdependencies of viewpoint beyond mere hierarchies and changing viewpoints over time and is hence future work for our model.

\section{Acknowledgements}
The first author was supported by the Leib\-niz-Science\-Campus Postdigital Participation funded by the Leibniz Association (Leibniz-Gemeinschaft).
We also like to thank our anonymous reviewers for their helpful comments and suggestions.
 
\bibliographystyle{elsarticle-num} 
\bibliography{dke_full_article}





\end{document}